# Ultrafast Magnetization Induced by Raman-Active Axial Chiral Phonons

**Authors:** Chuankun Huang[1†], Alexander Milner[2†], Jiaming Luo[1,3†], Jianbo Ye[1,3], Gaihua Ye[4], Junjie Zhang[1], Cynthia Nnokwe[4], Boris I. Yakobson[1], Rui He[4,5], Valery Milner[2*], Hanyu Zhu[1*]

**Affiliations:**

[1]Department of Materials Science and NanoEngineering, Rice University, Houston, TX 77005, U.S.A.

[2]Department of Physics & Astronomy, The University of British Columbia, V6T 2K9, Vancouver, Canada.

[3]Applied Physics Graduate Program, Rice University, Houston, Texas 77005, U.S.A.

[4]Department of Electrical and Computer Engineering, Texas Tech University, Lubbock, Texas 79409, U.S.A.

[5]Department of Physics, The University of Texas at Arlington, Arlington, Texas 76019, U.S.A.

†These authors contribute equally to the work.
*Corresponding authors. Email: Hanyu.Zhu@rice.edu, vmilner@phas.ubc.ca

**Abstract:**

Axial chiral phonons, which carry angular momentum and exhibit unusually large magnetic moments, provide a new degree of freedom for controlling the time-reversal symmetry and magnetic properties of quantum materials. While infrared-active axial phonons are efficiently excited to large amplitudes by circularly polarized terahertz pulses, their application may be constrained by short lifetimes, small penetration depths, low spatial resolution, and limited availability of optical sources. Here, we report that an axial Raman-active phonon mode in $CeF_3$ exhibits a significant magnetic moment and an exceedingly long lifetime that more closely matches the paramagnetic spin dynamics. The axial phonon population is resonantly driven by a near-infrared laser pulse with rotating linear polarization, known as an optical centrifuge, a coherent control scheme never applied to solids before. The phonon-driven magnetization observed by time-resolved Faraday rotation scales quadratically with incident power and rapidly decreases at high temperatures, consistent with the phonon inverse Faraday effect from many-body spin-phonon coupling. Our findings open a new avenue for using shaped laser pulses from widely accessible light sources to manipulate coherent axial chiral phonons and ultrafast spintronics.

**Main Text:**

Recently, chiral phonons involving circular motion of atoms in solids have attracted growing research interest as a new route for modifying material properties, including magnetization, optical activity, energy and information transport, and chemical reactivity (*1–7*). In particular, the large phonon-to-spin conversion, unexpected in a simple picture of ionic current-induced magnetic field, was experimentally observed across diverse materials and inspired numerous theoretical models (*8–18*). In non-magnetically ordered but structurally chiral materials, net atomic rotation may be induced by processes breaking time-reversal symmetry, including charge, heat, or light propagation (*19*, *20*). But in non-chiral materials, the angular momentum must be transferred from external stimuli. Many previous works have utilized terahertz (THz) or mid-infrared (IR) light to drive IR-active phonons at resonant frequencies, thereby controlling the angular momentum via circular polarization (CP) (*8*, *9*, *11*). The ultrafast lattice and spin dynamics are then analyzed by time-resolved spectroscopy. However, one challenge in understanding these light-driven magneto-optic responses is that IR-active phonons exhibit polaritonic behavior, which complicates the extraction of real atomic displacements as a function of frequency (*21*). Moreover, the resonant phonon Reststrahlen band often limits the chiral interaction near the surfaces, which are quite sensitive to experimental geometry, interfacial properties, or even the environment (*22*).

Alternatively, Raman-active phonons were also found to be strongly coupled with spins. In fact, magneto-Raman spectroscopy is much more widely available, leading to the first discovery of an anomalously large phonon Zeeman effect in $CeF_3$ in the 1970s and in many more materials recently (*15*, *16*, *23–29*). Coherent excitation of Raman-active phonons, via methods such as impulsive stimulated Raman scattering and coherent anti-Stokes Raman scattering, is widely used to study lattice dynamics in condensed matter physics and to identify vibrational signatures of compounds in chemistry and biology (*30–32*). Yet transferring optical angular momentum via an impulsive Raman process is not straightforward because the photons are absorbed and emitted into the same incoming light pulses. Using two ultrashort, time-delayed pulses with different linear polarizations, or two frequency-chirped pulses with opposite circular polarizations, rotational control has been demonstrated in crystalline solids and gas-phase molecular ensembles (*33*, *34*). The latter is more commonly known as an "optical centrifuge" for spinning molecules at extremely high rotational frequencies and inducing molecular magnetism (*35–39*). Applied to solids, this technique potentially enables unprecedented precision in controlling lattice dynamics and strongly driven atomic motion in the anharmonic regime.

Here, we demonstrate that chiral Raman-active phonons transiently magnetize paramagnetic $CeF_3$ when driven by an optical centrifuge, which is applied to solid-state systems for the first time. Surprisingly, this effect is not maximized for the modes known to exhibit the strongest spin-phonon coupling and the largest phonon magnetic moments. Instead, the magnetization arises in a low-frequency Raman mode at 2.35 THz that has previously been overlooked in the literature. The phononic Zeeman effect of this newly identified mode was verified by polarization-resolved, magnetic-field-dependent Raman spectroscopy. The phonon population lifetime of 281 ps, observed directly by time-resolved Raman scattering at 1.4 K, far exceeds the coherence lifetime deduced from its Raman linewidth. This lifetime is two orders of magnitude longer than that of other phonon modes and surpasses the paramagnetic spin relaxation time. Therefore, despite the moderate effective magnetic field, a more sizable spin response is observed when the centrifuge aligns with this phonon mode than with higher-frequency phonons, as measured experimentally by time-resolved Faraday rotation. This phonon-induced inverse Faraday effect (PIFE) arises only in the simultaneous presence of both light beams of the centrifuge, after carefully subtracting the trivial optical inverse Faraday effect (OIFE) from each individual beam, and scales quadratically with the incident pulse energy, as expected from stimulated Raman excitation. The sharp phonon resonance and its temperature dependence further rule out direct stimulated excitation of crystal electric field (CEF) transitions. Our findings bridge a gap in chiral phonon research, clearly separate the phononic and electronic origins of light-driven spin dynamics, and offer a more universal approach to modulating the magnetic states of quantum materials and spintronic devices.

First, we experimentally observed a pair of Raman-active phonon modes with unusual magnetic moments at a low frequency of 2.35 THz in $CeF_3$ crystals (hereafter labeled as $E_g^{(1)}$) via magneto-Raman spectroscopy (Fig. 1A). The axial phonons rotating in opposite directions are differentiated by incident left- and right-handed polarization (⟳ and ⟲), and the scattering signal is collected in the off-diagonal channel (i.e., opposite polarization rotation in the lab frame but same optical circular polarization in the reflection geometry). This $E_g^{(1)}$ mode was also verified in our first-principles calculation (Table S1) and in the Raman spectrum of $LaF_3$ (*40*), which has no 4f electrons, and thus cannot be of electronic origin.

The mode exhibits discernible Zeeman splitting, albeit much smaller than other $E_g$ modes (Fig. 1B) (*41*). The magnetic moment at 11 K is $0.07\mu_B$, still much larger than expected from nuclear rotation alone (*42*). What truly sets this mode apart is its narrow linewidth, which is close to the instrumental broadening. After deconvolution, the elastic scattering lifetime of the phonon is more than 6 ps, 10 times longer than the IR-active mode associated with strong chiral phonon-induced magnetization (*8*). Density functional perturbation theory (DFPT) reveals that the atomic displacement primarily involves the relative shear motion between the two $CeF_3$ layers in the unit cell (Fig. 1C and Fig. S2). Such motion has unusually low frequencies compared with typical optical phonons, thereby limiting the phase space for three-phonon scattering to acoustic phonons and extending the lifetime (*43*, *44*). The disadvantage of this mode is that $Ce^{3+}$ ions displace along with the $F^-$ ions, unlike in the IR-active $E_u$ modes, and thus may cause less modulation to the crystal fields with a weaker spin-phonon coupling and renormalization of phonon frequencies mediated by CEF interactions (*14*, *45*).

We then investigated the lattice and spin dynamics excited by the optical centrifuge setup (Fig. 2A). Femtosecond pulses from a Ti:sapphire laser system (35 fs Fourier transform limit, 1 kHz repetition rate, 796 nm central wavelength) are split into centrifuge and probe beams. The centrifuge field is a linearly polarized optical pulse whose polarization vector undergoes rotation about the propagation axis. Through the interaction with the induced dipole, the field exerts a torque on molecules, forcing them to follow the rotating polarization with controlled angular accelerations (up to 100 GHz/ps) and terminal frequencies (between 0 and 10 THz) (*35*, *46*). Here, we set the centrifuge pulse with a duration $\tau_{\mathrm{FWHM}} = 139$ ps to a tunable, nearly constant frequency of polarization rotation (cfCFG, Supplementary Sec. 3). The resulting field has a helical structure (Figs. 1C and 2A) and thus is truly chiral. Yet the chirality does not have physical significance here, so we only focus on the axial property of the phonon. For measurements of the oscillation of linear birefringence (Fig. S3), the shortest probe duration ~ 70 fs was used. The Raman and Faraday rotation probe pulses are shaped in a standard 4f pulse shaper (*47*), where their spectral bandwidth is narrowed to 0.03 THz, stretching them in time to about 5 ps.

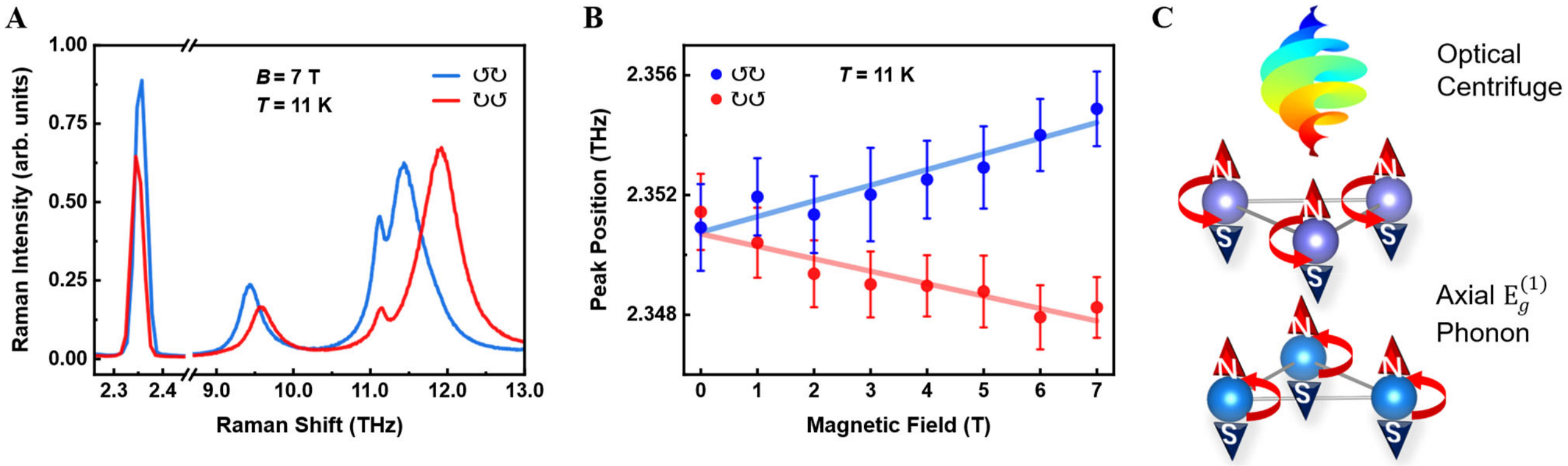

**Fig. 1. Magneto-Raman-active axial phonons in $CeF_3$. (A)** CP-resolved magneto-Raman spectrum of $CeF_3$ at 11 K, highlighting three pairs of field-dependent $E_g$ Raman modes that split under high magnetic field 7 T at 11 K. The symbols ↻ and ↺ denote clockwise and counterclockwise polarization rotation of the incident and scattered light in the lab frame. **(B)** The frequency splitting of the lowest degenerate $E_g^{(1)}$ mode as a function of the magnetic field at 11 K, indicating a significant magnetic moment and spin-phonon coupling. (**C**) Schematic of the $E_g^{(1)}$ axial phonon mode centered around 2.35 THz driven by an optical centrifuge. Two layers of $Ce^{3+}$ ions in triangular lattices are denoted in blue and purple. Their in-plane shear displacements are opposite but rotate in the same direction with a relative phase delay of $\pi$.

We found the prominent excitation resonance matching the $E_g^{(1)}$ frequency by time-resolved Raman scattering, after scanning the centrifuge frequencies in the range of 1–7 THz. The probe is circularly polarized and scatters into either Stokes or anti-Stokes Raman sideband, depending on whether its polarization rotates in the same or opposite direction to the centrifuge rotation (*34*). This property, stemming from the conservation of angular momentum, enables us to determine the axial directionality of the resonantly excited phonon modes (Fig. 2B). For the $E_g^{(1)}$ resonance, we observe that changing the direction of the centrifuge results in the corresponding change in the sign of the predominant Raman peak, indicating directionality in the phonon response to the centrifuge excitation. Although $CeF_3$ has a few more Raman-active modes within the range, we observed only one resonance, most likely because the scattering cross-sections of the other phonons and the CEF state are smaller, and their lifetimes are much shorter, which suppresses excitation efficiency under a long cfCFG driving pulse.

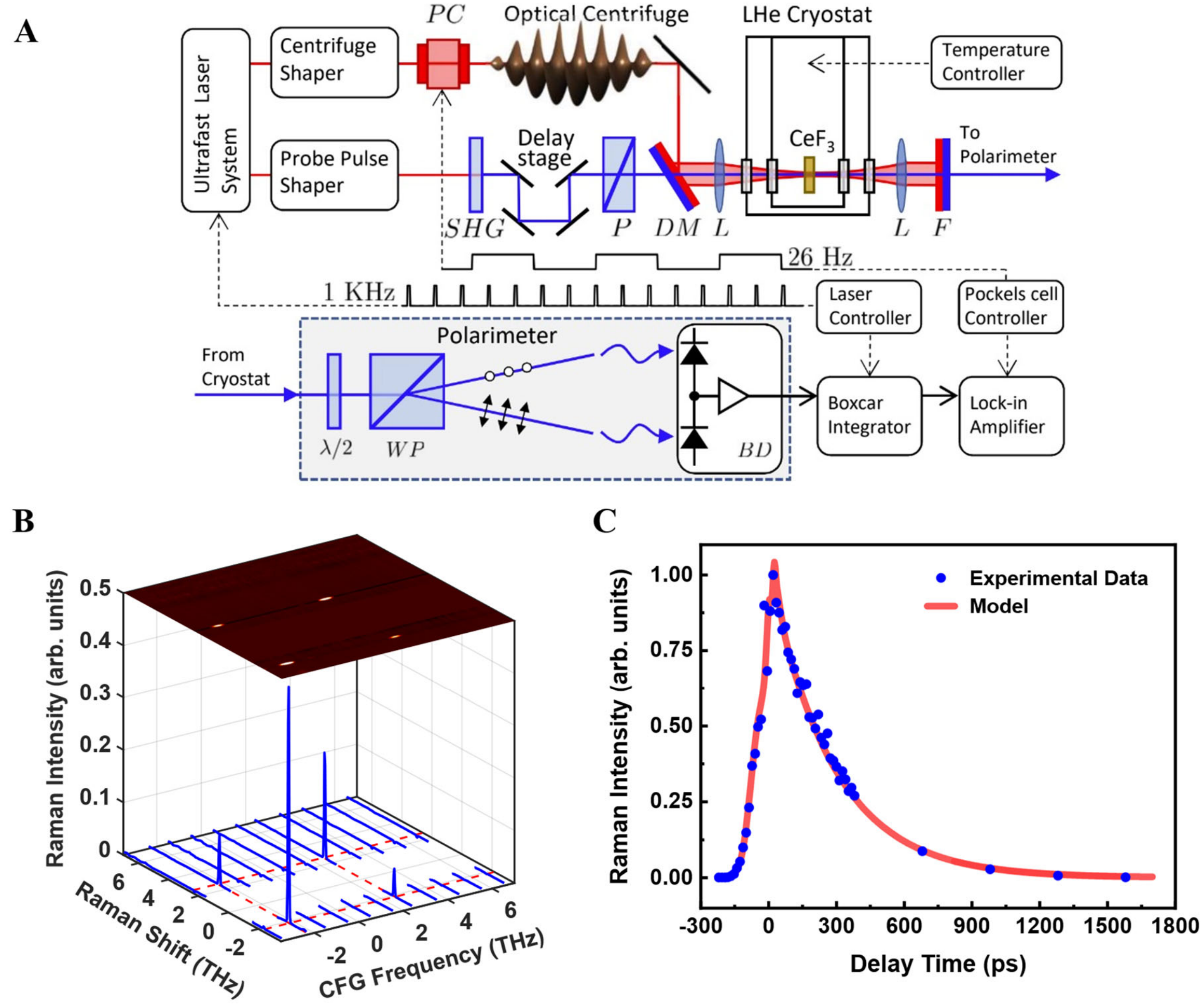

**Fig. 2. Centrifuge-field-driven $E_g^{(1)}$ phonon dynamics in $CeF_3$. (A)** The setup consists of centrifuge pulses centered at 796 nm, whose direction of rotation is modulated by a Pockels cell, and frequency-doubled probe pulses delayed relative to the pump. These two beams are combined in a collinear geometry and focused on a bulk $CeF_3$ crystal immersed in a liquid helium cryostat. After passing through the cryostat, probe pulses are spectrally filtered from the pump and sent either to a spectrometer or to a time-gated polarization analyzer. PC: Pockels cell; SHG: second harmonic generation crystal; P: polarizer; DM: dichroic mirror; L: lens; F: dichroic filter; $\lambda/2$: zero-order half-wave plates; WP: Wollaston prism; BD: balanced detector. **(B)** Experimentally observed Raman spectra as a function of centrifuge frequency, whose sign indicates the direction of the centrifuge rotation. Blue lines are selected cross-sections of the map, and red dashed lines label the phonon resonant frequency at ± 2.35 THz. **(C)** Resonantly driven time-resolved Raman intensity of $E_g^{(1)}$ phonon mode is fitted using a four-wave mixing model, yielding a population lifetime $\tau_{\mathrm{ph}} = 281 \pm 7$ ps.

An ultralong-lived phonon dynamics is illustrated by the Raman intensity in $CeF_3$ at 1.4 K (Fig. 2C). The coherent phonon field $Q$ is modeled as a mechanical oscillator driven by the optical field $E$ inside the material experienced by the phonons (*8*, *48*):

$$\frac{d^2Q}{dt^2} + \frac{1}{\tau_{\rm ph}}\frac{dQ}{dt} + \Omega^2 Q = \frac{\Omega}{\hbar}\sum_{ij}\frac{\partial \alpha_{ij}}{\partial Q}E_iE_j. \tag{1}$$

Here, $\Omega$ is the phonon's angular frequency, $\tau_{ph}$ is the phonon lifetime, $E_iE_j$ is the light field correlation in each spatial coordinate $i,\ j$, and $\frac{\partial \alpha_{ij}}{\partial Q}$ represents the phonon-induced polarizability change of materials. Fitting the experimental Raman signal $R(t)$, which contains the contributions from phonon population $n(t) = |Q(t)|^2$ and non-resonant four-wave mixing of the pump pulse (Supplementary Sec. 6), results in a phonon population lifetime $\tau_{\rm ph} = 281 \pm 7$ ps at 1.4 K. This $\tau_{\rm ph}$ is much longer than the linewidth-limited coherence lifetime resolvable in static Raman spectra, indicating that inhomogeneous broadening dominates the Raman scattering process. Importantly, the sign of the frequency shift of the dominant Raman peak in Fig. 2B changes together with the helicity of the centrifuge field, demonstrating that the excited phonons maintain angular momentum despite inhomogeneous dephasing. In this case, the local lattice anisotropy is weak compared to non-local inhomogeneity, such that the left- and right-CP $E_g^{(1)}$ phonons do not mix with each other (*49*). Since the effective magnetic field solely depends on the local phonon angular momentum and does not require global phase coherence, the spin accumulation is driven by the phonon effective magnetic field proportional to the incoherent Raman scattering intensity.

Next, we observed the magnetization dynamics $M(t)$ induced by axial $E_g^{(1)}$ using time-resolved polarimetry (Fig. 3A). Setting the probe linearly polarized, its polarization rotation after passing through the sample is measured using the standard balanced detection scheme (Fig. 2A). The excitation of $E_g^{(1)}$ phonons lead to an oscillatory linear birefringence signal (Fig. S3), while the spins cause a slowly evolving Faraday rotation. Therefore, stretching the probe pulses in the pulse shaper enables us to average out the birefringence oscillations (occurring with a period of 422 fs) and resolve the much weaker magnetization effect from spins, whose relaxation dynamics are modeled by paramagnetic relaxation

$$\frac{dM}{dt} = \frac{1}{\tau_{\rm spin}}\left[\frac{\chi B_{\rm eff}(t)}{\mu_0} - M\right], \tag{2}$$

where $\chi$ is the magnetic susceptibility of $CeF_3$ (*50*). We assume $B_{\rm eff}(t) = B_0 n_{\rm ph}(t)$ with the amplitude $B_0$ as a free parameter, where $n_{\rm ph}(t) = |Q(t)|^2$ is the normalized phonon population extracted from the Raman scattering intensity, unlike the previous work in which the field amplitude was fixed by theory (*8*). Another distinction is that $\tau_{\rm ph} > \tau_{\rm spin}$ prominently influences both the temporal evolution and the magnetization amplitude. The fitted spin lifetime is $74 \pm 5$ ps at 1.4 K, likely limited by dipolar interactions at high density. We also confirmed that the magnetization switches sign for centrifuges with opposite centrifuge helicities (blue and red dots in Fig. 3A).

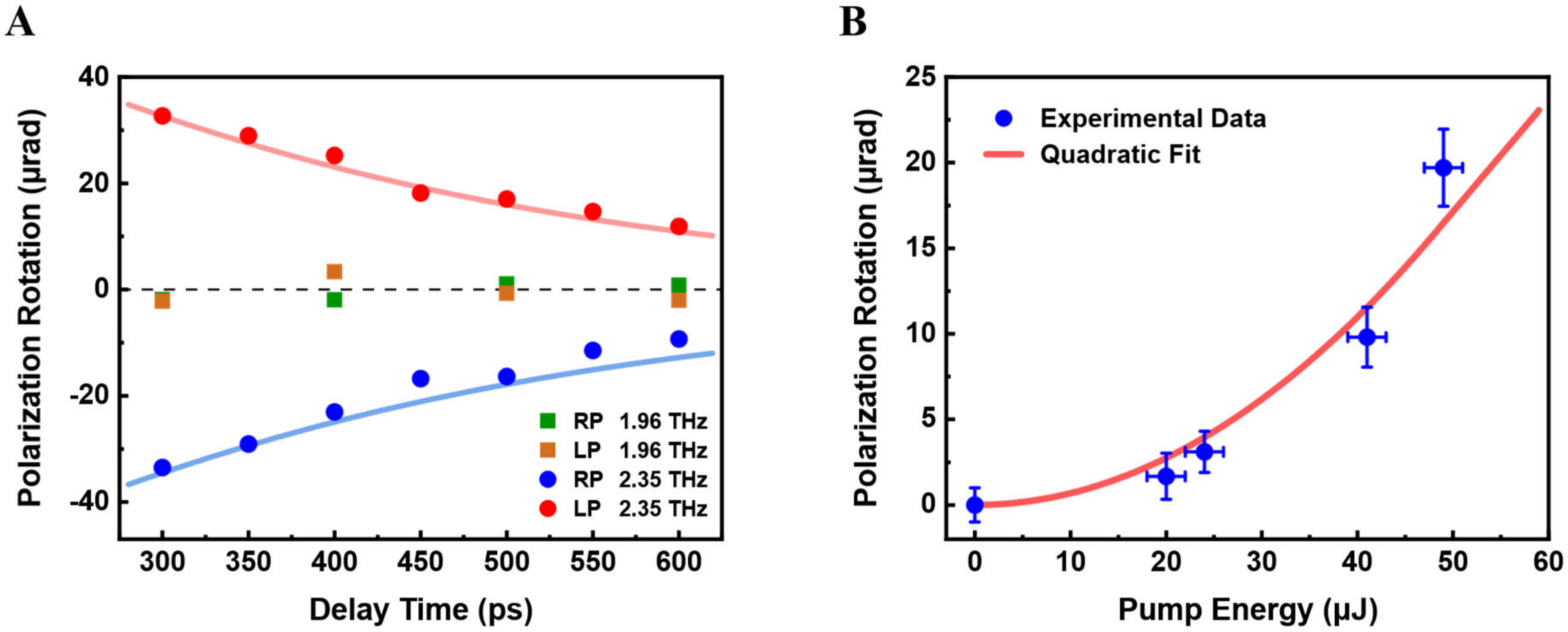

**Fig. 3. Ultrafast magnetization induced by axial Raman-active phonons.** **(A)** Magnetization dynamics measured by Faraday rotation beyond the pump pulse duration. When the excitation frequency is shifted from a phonon resonance at 2.35 THz (blue and red dots) to a non-resonant frequency of 1.96 THz (orange and green squares), the Faraday rotation vanishes, clearly demonstrating the role of phonon-mediated spin dynamics. RP/LP: right/left-circularly polarized centrifuge field; solid curves: fits from the spin relaxation model. **(B)** Faraday rotation observed at a fixed probe delay 350 ps at 1.4 K under resonant excitation, showing a quadratic dependence on pump energy expected from PIFE instead of OIFE.

Here, we rule out three major competing mechanisms by which strong optical excitations can cause a rotation of the probe polarization. First is the nonlinear Kerr effect from the pump field itself, which arises from residual dispersion and the breaking of Kleinman's symmetry relation, and prevents us from measuring the spin polarization during the pump pulses (*51*). All our spin dynamics were observed by delaying the probe at $2.5\tau_{\mathrm{FWHM}} = 350$ ps, when the pump intensity falls below 0.06% of peak power (for either Gaussian or $\mathrm{sech}^2$ shaped pulses), thus the nonlinear effects become negligible. Second is the mechanical Faraday effect, also known as "polarization drag", previously demonstrated in rotating dielectric solids and molecules, which induces an optical polarization rotation (*52–54*). This electrical polarizability effect, estimated at $10^{-8}$ rad in our experiment (Supplementary Sec. 6), is conceptually distinct from PIFE, in which the phonon exerts an effective magnetic field on spins, with the magneto-optic rotation as a secondary result. Third is OIFE from each individual CP component of the centrifuge. From the Faraday constant of $CeF_3$, the estimated OIFE from one CP component can be three orders of magnitude larger than PIFE (Supplementary Sec. 6). Fortunately, these two CP fields have nearly equal intensity, which causes opposite effective magnetic fields on the spins; and the Faraday rotation due to the remnant spins induced by OIFE falls below 1 µrad at the probe delay much larger than the spin lifetime, as shown at non-resonant driving frequencies (square dots, Fig. 3A). Furthermore, the magnetization at resonant driving frequencies is a quadratic function of the pump intensity, i.e., proportional to the phonon populations (Fig. 3B), contrasting a linear power dependence of remnant OIFE.

We confirm that the phonon magnetic moment originates from many-body CEF-lattice coupling as opposed to a direct Raman excitation of electronic transitions (*14*). Previously, a simple microscopic theory explained that the phonon magnetic moment arises from its coupling to the CEF transitions of many electrons (*55*). Although this coupling means the phonon-like resonance always contains a small CEF component, the Zeeman splitting of the phonons is not caused by the magnetic moment of the CEF component, but by changes in spin population. To illustrate this crucial point, we note that the hybrid wavefunction is independent of temperature $T$ when we neglect thermal excitation of high CEF levels, but the paramagnetic spin population in a magnetic field $B$ shifts as $\Delta n \propto \tanh(\frac{g\mu_B B}{k_B T})$, where $g$ is the known out-of-plane $g$-factor of a single $Ce^{3+}$ spin (*56*). The temperature-dependent phonon Zeeman splitting at $B = 7$ T is shown in Fig. 4A, giving saturation splitting $\Delta\Omega = 0.008$ THz when the electrons are fully spin-polarized. The strong inverse temperature dependence verifies that, for the $E_g^{(1)}$ mode, the many-body spin-lattice coupling dominates over the small electronic wavefunction contribution to the magnetic response.

Likewise, the inverse process of the phonon effective magnetic field is not caused by OIFE through the CEF component of the phonon mode, but by PIFE through the presence of axial phonon population. First, a direct OIFE through the virtual excitation of CEF by an off-resonant stimulated Raman process is proportional to $\frac{1}{(\omega-\omega_0)^2}$, where $\omega_0$ is the CEF frequency, and $\omega$ is the frequency of the optical centrifuge (*57*). This term can be ruled out easily due to its flat spectrum at large detuning. Second, the Raman excitation of a phonon mode hybridized with CEF generates an OIFE proportional to the instantaneous phonon field and the fraction of the CEF component $\frac{1}{\tau_{ph}^2(\omega-\Omega_{ph})^2+1}\frac{g^2}{(\Omega_{ph}-\omega_0)^2}$, where $g$ is the effective coupling constant between the phonon and CEF. This term would exhibit a sharp resonance at the phonon frequency $\Omega_{ph}$ as observed in the experiment. It could be important when the phonon and CEF are nearly resonant ($g \sim |\Omega_{ph} - \omega_0|$), which is not the case for $E_g^{(1)}$ given the relatively small $\Delta\Omega$ compared with some high-frequency phonon modes (*8*, *23*). Third, the PIFE through the effective magnetic field of phonons is proportional to the coherent spin-phonon coupling constant and duration of the phonon field $\frac{g^2\tau_{ph}}{\Omega-\omega_0}$. The key difference in physics underlying the two contributions is that OIFE requires electronic coherence, whereas PIFE requires only phonon coherence, which scales favorably at large detuning $|\Omega - \omega_0|$ and can persist much longer after the optical pulses. Because of such a distinction in dynamics, one should not naively think of phonons as merely a resonance booster of the electronic optical response. Instead, angular momentum-conserving axial phonons indeed exhibit non-trivial magnetic properties different from regular phonons and the electrons that couple to them.

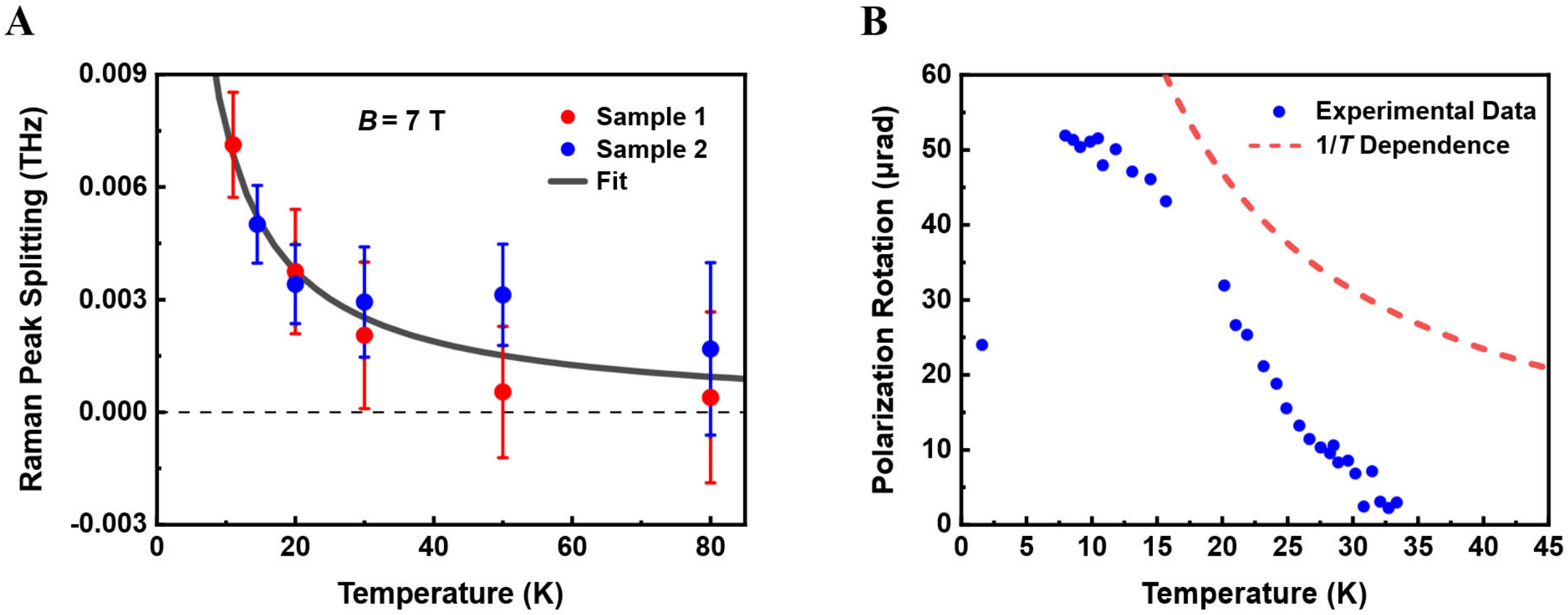

**Fig. 4. Temperature dependence of the spin-phonon coupling in $CeF_3$. (A)** Phonon Zeeman splitting at a fixed magnetic field 7 T closely follows the spin population imbalance $\Delta n \propto 1/T$ (solid curve), confirming that the phonon magnetic moment of $E_g^{(1)}$ originates from the many-body effect. **(B)** Phonon-induced magnetization (blue dots) measured at a fixed probe delay 350 ps is compared to a $1/T$ dependence estimated from a quasi-static model (red dashed line). The decreasing magnetization at higher temperatures is attributed to a shorter phonon lifetime. The reduced magnetization at 1.4 K may be caused by slow spin excitation. The 2–8 K gap is due to the unreliable temperature control from helium convection.

Figure 4B further distinguishes the PIFE, OIFE, and mechanical Faraday mechanisms by measuring the resonantly driven polarization rotation as a function of temperature $T$ at a given probe delay 350 ps. In a simple OIFE or quasi-static PIFE picture without any phonon or spin dynamics, $\theta_F \propto V(T) \propto 1/T$ (red dashed line and Supplementary Sec. 6), which deviates from observation at both high and low temperatures. Between 10–25 K, when the phonon lifetime remains long, quasi-static PIFE predicts the correct order of magnitude. Above 25 K, the magnetization decreases much more quickly because in PIFE the optical phonon decay rate increases with the acoustic phonon population (*8*, *44*), in contrast to OIFE. Surprisingly, $M(T)$ also decreases below 10 K. The deviation indicates slower spin excitation by the effective magnetic field and contradicts the mechanical Faraday picture. It is partially corroborated by the longer spin relaxation lifetime observed at 1.4 K compared to that measured at higher base temperatures (*8*). But it is also likely that the spin lifetime is not a constant but a function of Zeeman splitting at very low temperatures, such that the excitation process under an external field is less efficient than the relaxation process under no external field. From a practical perspective, an intermediate temperature at which the spin and phonon lifetimes match the desired control pulse duration is ideal for efficient ultrafast spintronics.

Finally, we discuss possible routes towards a stronger phonon-induced spin response. From Eq. (1), under a peak optical intensity of ~ 47 GW/cm$^2$, we would reach a peak coherent phonon field $Q \simeq 0.7$, corresponding to a maximum atomic displacement of 2–3 pm, which is quite substantial in solids. The expected field strength should be on the order of 20 mT and may further increase with incident power. Since $CeF_3$ has a large band gap of over 4.5 eV, a longer-wavelength excitation can significantly increase the multi-photon absorption threshold, corresponding to an effective field on the tesla scale. However, the observed effective field strength is on the order of 10 μT at 10 K. Such a discrepancy may come from four sources. First and foremost, inhomogeneous broadening means coherent phonons are only in resonance with an ideal pump pulse in a small fraction of regions. Second, reaching the theoretical phonon amplitude requires an optical centrifuge that maintains its resonant frequency with ~ 1 GHz precision. In practice, due to the third-order dispersion of our infrared pulses, the centrifuge frequency varies by up to 100 GHz over the centrifuge pulse (*58*). These two effects reduce the excitation efficiency by 2 orders of magnitude (Fig. S4A). Hence, the excitation efficiency of axial phonons may be improved by higher crystalline quality, a tighter focus to reduce sample volume, and dispersion correction in the centrifuge. Third, the Raman tensor was calculated at zero frequency and did not account for any optical transitions (the static Placzek approximation). As a result, a deviation by an order of magnitude in Raman excitation efficiency is possible between experiment and theory (*59*). Fourth, the low-temperature Verdet constant of $CeF_3$ was extrapolated from an empirical relationship with magnetic susceptibility with possible errors. More accurate computation and characterization will potentially bridge the gap in the future.

In summary, we discovered long-lived, Raman-active, axial chiral phonons with strong spin-phonon coupling in single crystalline $CeF_3$ using static magneto- and time-resolved Raman spectroscopy. The atomic rotation at 2.35 THz was coherently driven by an optical centrifuge and induced a magnetic response whose direction is controlled by the polarization rotation of the centrifuge. The time-resolved lattice and spin dynamics are explained using semiclassical rate equations and compared with the many-body coupling model, revealing possible directions for drastically boosting spin-manipulation efficiency. The successful optical control of Raman-active chiral phonons makes chiral manipulation of quantum materials more accessible in diverse materials and laboratory settings and potentially enables ultrafast optical operation of nanoscale spintronic devices.